\documentclass{WileyChemistry-template}

\usepackage{changes}

\title{\raggedright 
Spin Excitations of High Spin Iron(II) in Metal-Organic Chains on Metal and Superconductor
}

\author{
\begin{minipage}{\textwidth}
	Jung-Ching Liu,*\textsuperscript{,[a]} Chao Li,\textsuperscript{[a]} Outhmane Chahib,\textsuperscript{[a]} Xing Wang,\textsuperscript{[b],[c]} Simon Rothenb\"uhler,\textsuperscript{[b]} Robert H\"aner,\textsuperscript{[b]} Silvio Decurtins,\textsuperscript{[b]} Ulrich Aschauer,\textsuperscript{[b],[d]} Shi-Xia Liu,*\textsuperscript{,[b]} Ernst Meyer,\textsuperscript{[a]} R\'emy Pawlak*\textsuperscript{[a]}
\end{minipage}
}

\newcommand{\affiliation}{
\begin{itemize}

\item[{[a]}] Dr. J.-C. Liu, Dr. C. Li, O. Chahib, Prof. E. Meyer, Dr. R. Pawlak\\
Department of Physics, University of Basel, Klingelbergstrasse 82, 4056 Basel, Switzerland\\
E-mail: jungching.liu@unibas.ch, remy.pawlak@unibas.ch

\item[{[b]}] Dr. X. Wang, Dr. S. Rothenbühler, Prof. R. Häner, Prof. Dr. S. Decurtins, Prof. U. Aschauer, PD Dr. S.-X. Liu*\\
Department of Chemistry, Biochemistry and Pharmaceutical Sciences, W. Inäbnit Laboratory for Molecular Quantum Materials, University of Bern, Freiestrasse 3, 3012 Bern, Switzerland\\
Email: shi-xia.liu@unibe.ch

\item[{[c]}] Dr. X. Wang\\
Paul Scherrer Institut, Forschungsstrasse 111, 5232 Villigen PSI, Switzerland

\item[{[d]}] Prof. U. Aschauer\\
Department of Chemistry and Physics of Materials, Paris-Lodron University Salzburg, Jakob-Haringer-Strasse 2a, 5020 Salzburg, Austria

\end{itemize}
}

\renewcommand{\abstract}{
Many-body interactions in metal-organic frameworks are fundamental for emergent quantum physics. Unlike their solution counterpart, magnetization at surfaces in low-dimensional analogues is strongly influenced by magnetic anisotropy induced by the substrate and still not well understood.
Here, we use on-surface coordination chemistry to synthesize on Ag(111) and superconducting Pb(111) an iron-based spin chain by using pyrene-4,5,9,10-tetraone precursors as ligands. Using low-temperature scanning probe microscopy, we compare their structures and low-energy spin excitations of coordinated Fe atoms with high $S$ = 2 spin-state. Although the chain and coordination centers are identical on both substrates, the long-range spin-spin coupling due to a superexchange through the ligand observed on Ag is absent on Pb(111). We ascribe this reduction of spin-spin interactions on Pb to the depletion of electronic states around the Fermi level in the Pb(111) superconductor as compared to silver.}

\newcommand{\keywords}{
	coordination chemistry \textbullet\ 
	magnetocrystalline anisotropy \textbullet\ 
	scanning probe microscopy \textbullet\ 
	spin excitation \textbullet\ 
	superconductivity
}

\begin{document}

\twocolumn[\vspace{-1.5cm}\maketitle\vspace{-1cm}
	\textit{\dedication}\vspace{0.4cm}]
\small{\begin{shaded}
		\noindent\abstract
	\end{shaded}
}

\begin{figure} [!b]
\begin{minipage}[t]{\columnwidth}{\rule{\columnwidth}{1pt}\footnotesize{\textsf{\affiliation}}}\end{minipage}
\end{figure}

Many-body interactions in metal organic frameworks (MOFs) are at the forefront of emergent quantum physics such as superconductivity,\cite{Zhang2017,Takenaka2021} ferromagnetism\cite{Li2017,Park2021,Lobo-Checa2024,Pitcairn2024} or quantum spin liquid.\cite{Yamada2017,Misumi2020} With their high versatility, two-dimensional metal-organic frameworks (2D MOFs) are prime candidates for inducing such exotic phenomena\cite{Wang2013} since they use coordination chemistry to associate atomic magnets with multi-dentate organic ligands at surfaces through self-assembly processes.\cite{Stepanow2008,Carbone2011,Dong2016} 2D MOFs thus allow the precise arrangement of interacting spins into complex lattices and a control over their exchange interaction by judiciously adjusting the inter-adatom distance. Lattice geometry and adatom spacing over the surface are controlled by the rational design of the organic linkers,\cite{Stepanow2008} and their coordination to metal centers, which preferentially show high spin magnetic moments.\cite{Gambardella2009,Abdurakhmanova2013}

In MOFs, magnetic exchange interactions are interpreted in terms of superexchange interactions through the organic ligands.\cite{Xu2020,Umbach2012,Abdurakhmanova2013} At surfaces, surface-induced magnetic anisotropy (MA) forces individual magnetic moments in 2D MOFs to orient along preferential directions, while electronic overlap between these local moments through the metal electrons in the regime of the Ruderman-Kittel-Kasuya-Yosida (RKKY) exchange\cite{Ruderman1954,Kasuya1956,Yosida1957} can also lead to complex magnetic phases.\cite{Umbach2012} These effects can be seen as an opportunity to tune the 2D MOF magnetism  not only by considering magnetic anisotropy\cite{Gambardella2009,Mallada2021} and interatomic coupling\cite{Li2023a} but also the intrinsic band structure of the underlying substrate. For instance, superconductors have shown to protect excited spin states\cite{Heinrich2013a,Vaxevani2022} due to opening of the energy gap $\Delta$ around the Fermi level $E_{\textrm{F}}$ when the condensation of electrons into Cooper pairs occurs at low temperatures.  RKKY interaction between magnetic atoms on superconductors\cite{Liebhaber2022} has also been shown to play a pivotal role for the experimental realization of topological superconductivity.\cite{Pyhnen2014,Soldini2023,Jaeck2021} In this context, extending the synthesis of 2D MOFs on superconductors\cite{Ahmadi2016,Yan2021} is of paramount importance in engineering designer quantum materials, topological superconductors or protecting excited spin states in spin logic devices.

\begin{figure*}[t!]
\begin{center}
\includegraphics[width=15.5cm]{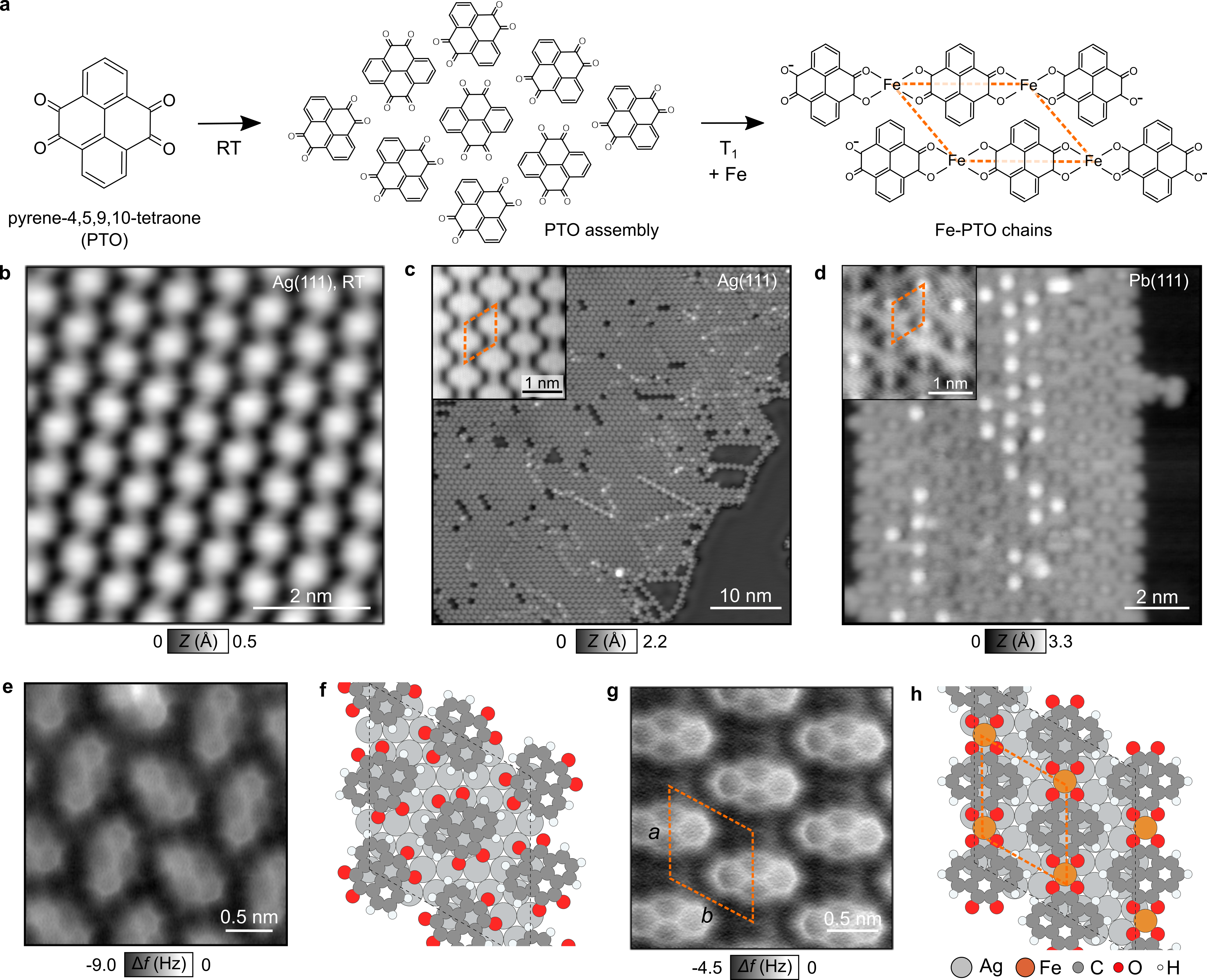}
\caption{Hierarchical synthesis of the Fe-PTO spin chain. 
\textbf{a)} Scheme of the reaction pathway using pyrene-4,5,9,10-tetraone precursor. Molecules self-assemble into a supramolecular network stabilized by hydrogen bonds. Annealing the substrate to $T_1$  $\approx$ 120-150 $^{\circ}$C while depositing Fe adatoms leads to the formation of the Fe-PTO chains. 
\textbf{b)} STM image of the PTO supramolecular network on Ag(111), ($I_{\textrm{t}}$ = 1 pA, $V_{\textrm{s}}$ = 500 mV).
\textbf{c)} STM image of the Fe-PTO chains on Ag(111) ($I_{\textrm{t}}$ = 1 pA, $V_{\textrm{s}}$ = 200 mV). The inset shows a close-up STM image of the structure. 
\textbf{d)} STM image of the Fe-PTO structure on Pb(111),($I_{\textrm{t}}$ = 80 pA, $V_{\textrm{s}}$ = 600 mV, inset: $I_{\textrm{t}}$ = 100 pA, $V_{\textrm{s}}$ = 50 mV).
\textbf{e)} AFM image with a CO-terminated tip of the PTO assembly on Ag(111), ($f_{0}$= 26.2 kHz, $A$ = 50 pm).
\textbf{f)} Relaxed structure calculated by DFT of the PTO assembly in registry with the Ag(111). 
\textbf{g)} AFM image of the Fe-PTO chains. The orange dashed line shows the Fe-Fe arrangement in the metal-organic network with lattice parameters $a$ = 8.7 \AA~and $b$ = 9.0 \AA.
\textbf{h)} Relaxed structure of Fe-PTO chain on Ag(111) obtained by DFT calculations.}
\label{fig1}
\end{center}
\end{figure*}
To characterize spin excitations at the atomic level, scanning tunneling microscopy (STM) and spectroscopy (STS) are particularly suitable techniques since they allow to probe the local density of states (LDOS) and spin states between tip and sample with high lateral and spectral resolution.\cite{Ternes2015} In an inelastic electron tunneling spectroscopy (IETS), tunneling electrons sense spin-flip excitations of a magnetic impurity coupled to  the itinerant electrons of a metal,\cite{Heinrich2004,Hirjibehedin2007} 
which manifest themselves in differential conductance (d$I$/d$V$) spectra as pairs of steps symmetric to the Fermi level $E_{\textrm{F}}$. This can be rationalized by a phenomenological spin Hamiltonian:\cite{Gatteschi2006,Jacob2018}

\begin{equation}
\label{eq1}
			\mathcal{H} = D\hat{S}_z^2 + E(\hat{S}_x^2 - \hat{S}_y^2),
\end{equation}

\noindent where $D$ and $E$ are the axial and transverse anisotropy parameters and $\hat{S}_{x,y,z}$ are  the three components of the spin operator along the corresponding axes. When magnetic moments are in close vicinity, spins can interact via orbital overlapping\cite{Hirjibehedin2006,Choi2016}, through RKKY interactions,\cite{Girovsky2017} or superexchange interaction, which can lead to collective spin eigenstates and additional spin-flip excitations at higher energies. 

Here, we compare the synthesis and spin excitations of a prototypical iron based metal-organic spin chain on Ag(111) and superconducting Pb(111) using pyrene-4,5,9,10-tetraone (PTO) molecules as ligands.\cite{DellaPia2016} Using STM and AFM,  we have found that the metal-organic structures are identical on both substrates. STS measured at 1 K on Fe sites reveals a series of low-energy steps attributed to spin excitations of the iron magnetic moment. On Ag(111), we observe steps at $\pm$ 20 meV but not on Pb(111), which are attributed to long range spin-spin coupling between the irons of the metal-organic chain. We attribute the absence of spin-spin coupling in the chains on Pb(111) to the depletion of electronic states around the Fermi level. 

The reaction pathway of Fe adatoms and PTO precursors into spin chains is shown in Figure~\ref{fig1}a. Experimentally, PTO molecules are sublimed onto the substrate kept at room temperature, leading to extended 2D PTO assemblies stabilized by hydrogen bonds (Figure~\ref{fig1}b, see Methods). In a second step, Fe adatoms are deposited while annealing  the substrate hosting the supramolecular assemblies to $T_{\textrm{1}}$ = 120-150$^{\circ}$C. This induces the PTO-Fe coordination and the formation of the spin chain (Figures~\ref{fig1}c and \ref{fig1}d). The oxidative potential of the bridging ligand stabilizes the Fe(II) dication.\cite{Skomski2015}

\begin{figure*}[t!]
\begin{center}
\includegraphics[width=16.5cm]{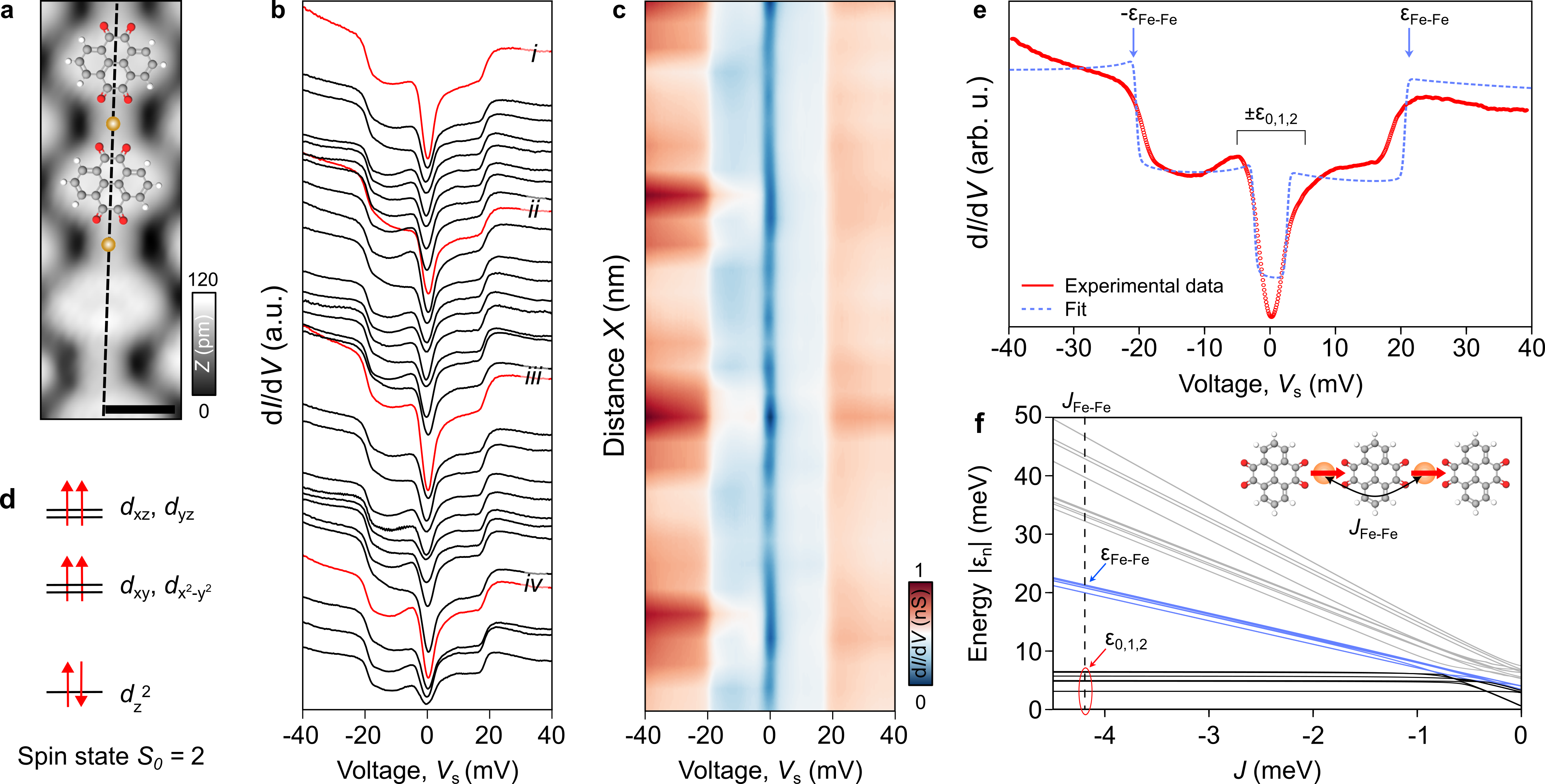}
\caption{Spin excitations of the Fe-PTO chain on Ag(111). 
\textbf{a)} STM image of a PTO-Fe chain ($I_{\textrm{t}}$ = 100 pA, $V_{\textrm{s}}$= 50 mV). Scale bar: 500 pm. 
\textbf{b)} Series of d$I$/d$V$ spectra recorded along the Fe-PTO chain of \textbf{a)}. Spectra measured around Fe sites are colored in red ($I_{\textrm{t}}$ = 300 pA, $V_{\textrm{s}}$ = 35 mV, $A_{\textrm{mod}}$ = 1 mV). 
\textbf{c)} d$I$/d$V$(V$_{\textrm{s}}$,X) cross-section of the datatset shown in \textbf{b}. \textbf{d)} Spin ground state of Fe calculated by DFT. It shows a spin ground state $S$= 2 and the $3d$ energy levels. 
\textbf{e)} Experimental d$I$/d$V$ spectrum of Fe (red curve) and the simulated curve (blue dotted line) considering two ferromagnetically coupled Fe spins (see inset of \textbf{f)}) ($I_{\textrm{t}}$ = 1.5 nA, $V_{\textrm{s}}$= 35 mV). The simulated curve shows a dip around $E_{\textrm{F}}$ denoted $\pm\varepsilon_{\textrm{0,1,2}}$ and a pair of steps at $\pm$ 20 meV ($\pm\varepsilon_{\textrm{Fe-Fe}}$) marked by a blue arrow.
\textbf{f)} Spin excitation energies as a function of the spin-spin exchange energy $J$. The magnetic exchange energy corresponding to the Fe-PTO simulated curve is shown with a dashed line  along $J$ = $J_{\textrm{Fe-Fe}}$). }
\label{fig3}
\end{center}
\end{figure*} 
Using AFM imaging with CO-terminated tips,\cite{Gross2011} we elucidate the chemical structure of the PTO assembly and the Fe-PTO spin-chains on Ag(111) (Figures~\ref{fig1}e and g). PTO molecules first self-assemble into an hexagonal arrangement of molecules rotated by 100$^{\circ}$ with respect to each other, the structure of which is confirmed by density functional theory (DFT) calculations (Figures~\ref{fig1}f). The supramolecular network, in registry with the Ag(111) lattice, is stabilized by intermolecular hydrogen bonds between neighboring ketone side groups and peripheral hydrogen atoms (C=O$\cdots$H-C).\cite{Pawlak2009} 

In Figure~\ref{fig1}g, the Fe coordination to oxygen atoms along the Fe-PTO chains is unambiguously resolved. The bonding geometry indicates a coordination number of four for each Fe atom as it is bound to four oxygen atoms from two neighboring PTO molecules. For completeness, we also compared the relaxed structures by DFT of Ag- and Fe-coordinated PTO networks and found that the Fe-coordinated system is 5.40 eV more energetically favorable than its Ag counterpart. The Fe-PTO molecular network is found in registry with the Ag surface with lattice parameters $a$ = 8.7 \AA~and $b$ = 9.0 \AA~as shown with orange dashed lines in Figure~\ref{fig1}g. DFT calculations further indicate a $S$ = 2 high-spin state of the Fe$^{\textrm{2+}}$ complex (Figures~\ref{fig3}d and Figure S1b) resulting from the coordination with the PTO molecules~\cite{Gambardella2009,Umbach2012}  We also acquired constant-height d\textit{I}/d\textit{V} maps at e$V_{\textrm{s}}$ $\pm$ 1.8 meV  and $\pm$ 20 meV (SI Figure S5c). At both energies, a homogeneous DOS distribution is observed at PTO molecules with a  slightly darker contrast at Fe sites which reflects the spin excitation dips of the corresponding d$I$/d$V$ spectra.

To experimentally confirm the Fe spin state, we performed a series of d\textit{I}/d\textit{V} spectra taken along a Fe-PTO chain (dashed line of Figure~\ref{fig3}a). At Fe sites, single-point d\textit{I}/d\textit{V} spectra colored in red in Figure~\ref{fig3}b systematically show a pronounced dip around $E_{\textrm{F}}$ marked $\pm\varepsilon_{\textrm{0,1,2}}$ accompanied by one pair of symmetric steps at about $\pm$ 20 meV ($\pm\varepsilon_{\textrm{Fe-Fe}}$). The dip is assigned to low-energy spin-flip excitations made possible by the magnetic anisotropy of the Fe atom in contact with the Ag(111) substrate, while steps at $\pm$ 20 meV are attributed to collective excitations between neighboring Fe spins. The d\textit{I}/d\textit{V} lineshape is consistently reproduced on molecule sites, which suggests the delocalization of these spin-flip processes across the entire Fe-PTO chains. We assume that, instead of a direct overlap of Fe $d$-orbitals, a long-range magnetic coupling between Fe atoms is the consequence of such spin-spin coupling. 

A representative d\textit{I}/d\textit{V} spectrum acquired at Fe sites is shown in red in Figure~\ref{fig3}e. To rationalize this, we first simulated the spectrum using the perturbative model developed by Ternes (see Methods) considering a single magnetic impurities of spin state $S$ = 2. The best agreement between simulation and experiment is only obtained for the low-energy spin excitations (i.e. $\lvert eV_{\textrm{s}} \rvert$ = $\varepsilon_{\textrm{0,1,2}}$ $\leq$ 10 meV) using the parameters $g$ = 2.1, $D$ = -0.8~meV and $E$ = 0.25~meV, while taking into account the substrate electron bath (i.e. potential scattering of $U$ = 0.2~meV and Kondo scattering of $J_K\rho_s$= -0.09~meV). As both axial and transverse anisotropy are considered in simulations, five spin states can be constructed using the single Fe impurity model. Considering the selection rule, only excitations with $\Delta m_z$ = $\pm$1 or 0 are possible. The ground state $\Psi_0$ has the largest weight in $|m_z\rangle$= $|\pm 2\rangle$ allowing the transition to $\Psi_1$ ($\Delta m_z$= 0), $\Psi_2$ and $\Psi_3$ ($\Delta m_z$= $\pm$1). The resulting low-energy spin excitations from simulation are then $\varepsilon_0$ = $\pm$0.2 meV, $\varepsilon_1$ =  $\pm$1.9 meV, and $\varepsilon_2$ =  $\pm$3.3 meV, respectively. In Figure~\ref{fig3}e, we marked the region of these spin excitations energies $\varepsilon_{\textrm{0,1,2}}$ which are responsible for the dip of the experimental spectra. Note also that to distinguish these low excitation energies of single atoms usually requires the use of a decoupling layer to reduce the DOS contribution from the normal metal substrate.\cite{Hirjibehedin2007,Otte2008,Choi2009} In our system, we assume that such decoupling from the Ag(111) is favored by the coordination with surrounding PTO molecules.\cite{Gambardella2009,Umbach2012}

\begin{figure*}[t!]
\begin{center}
\includegraphics[width=16.5cm]{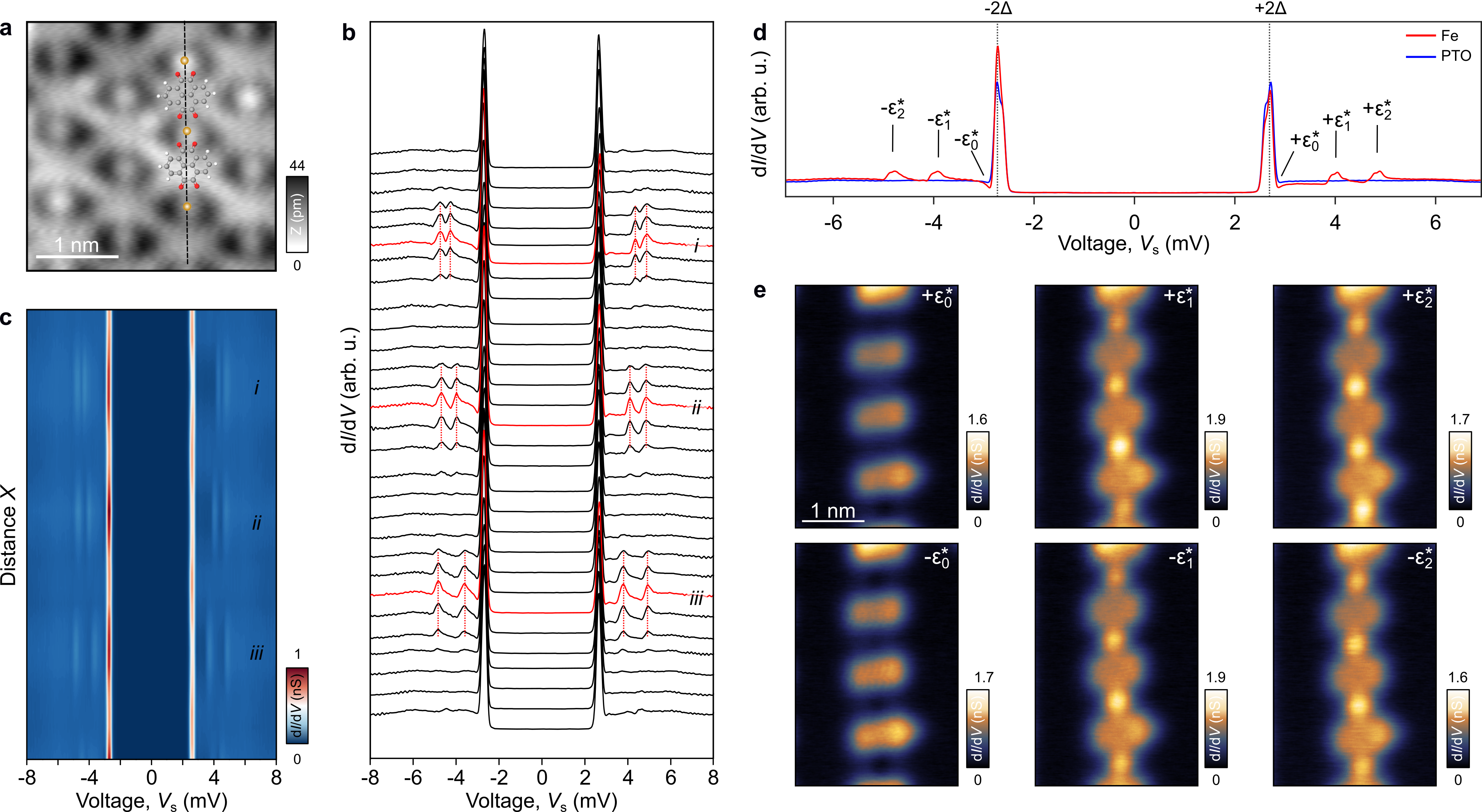}
\caption{Spin excitations of the Fe-PTO chain on Pb(111). 
\textbf{a)} STM image of the Fe-PTO chain on Pb(111), ($I_{\textrm{t}}$= 80 pA, $V_{\textrm{s}}$ = 9 mV). 
\textbf{b)} Series of d$I$/d$V$ spectra recorded along the black line of \textbf{a)}. d$I$/d$V$ spectra corresponding to Fe sites are plotted in red ($I_{\textrm{t}}$ = 300 pA, $V$ = 8 mV, $A_{\textrm{mod}}$ = 50 $\mu$V). 
\textbf{c)} d$I$/d$V$ cross-section showing asymmetric coherence peaks and resonances outside the gap attributed to spin excitations of the Fe sites marked $i$, $ii$ and $iii$, respectively. These features are absent on PTO molecules.
\textbf{d)} Representative d$I$/d$V$ spectra at Fe site (red) as compared to PTO molecule (blue), ($I_{\textrm{t}}$ = 800 pA, $V_{\textrm{s}}$ = 7 mV, $A_{\textrm{mod}}$ = 18 $\mu$V).
\textbf{e)} d$I$/d$V$ maps recorded at the spin-excitation energy of $\pm\varepsilon_{0}^*$, $\pm\varepsilon_{1}^*$ and $\pm\varepsilon_{2}^*$, respectively).  }
\label{fig6}
\end{center}
\end{figure*}
With the exception of the $\varepsilon_{\textrm{0,1,2}}$ spin excitations, the simulation using a single impurity model is unable to reproduce the $\varepsilon_{\textrm{Fe-Fe}}$ steps at $\pm$20~meV  (blue arrow of Figure~\ref{fig3}e). Inspired by spin-ladder compounds\cite{Silva2021} and other 2D MOFs on surfaces,\cite{Umbach2012} we next considered a chain of coupled spins with magnetic exchange interaction $J$ as depicted in the inset of Figure~\ref{fig3}f. According to DFT calculations (SI Figure S2), a ferromagnetic coupling is slightly favored as compared to an antiferromagnetic one. The spin Hamiltonian now writes as:

\begin{equation}
\label{eq2}
			\mathcal{H} = \sum_{i}^{2}\mathcal{H}_{i} + J(\vec{S_{1}}\cdot\vec{S_{2}}),
			\end{equation}
\noindent where $\mathcal{H}_{i}$ is the Hamiltonian of a single Fe. $J_{\textrm{Fe-Fe}}$ corresponds to the magnetic exchange coupling between the Fe spins along the Fe-PTO chain. Using $J_{\textrm{Fe-Fe}}$= -4.1~meV, the simulated spectra (blue dotted line of Figure~\ref{fig3}e) now features kinks at $\pm$20~meV providing the best agreement between theory and experiment. Figure~\ref{fig3}f last shows the evolution of these spin excitation energies as a function of $J$. The low-energy spin excitations $\varepsilon_{\textrm{0,1,2}}$ arise from the magnetic anisotropy of the Fe atom adsorbed on Ag(111) and remain constant and below 10 meV as $J$ increases. They are responsible for the dip centered to $E_{\textrm{F}}$ observed in the d$I$/d$V$ spectra. In addition, a series of spin excitations are observed at higher energy ($\geq$ 10 meV), which  linearly increases with $J$. From this plot, we thus estimate the magnetic exchange interaction $J_{\textrm{Fe-Fe}}$= -4.1~meV marked by a dashed line in Figure~\ref{fig3}f as it reproduces the $\varepsilon_{\textrm{0,1,2}}$ dip and the steps $\varepsilon_{\textrm{Fe-Fe}}$ at  $\lvert \textrm{e}V_{\textrm{s}}\rvert$ = 20 meV. 

Using the same preparation procedure, we next synthesize the Fe-PTO chains on Pb(111). The STM image Figure~\ref{fig6}a shows the lattice with parameters $a$ = 8.8~\AA~and $b$ = 9.1 \AA~in agreement with the structure on Ag(111), where iron atoms are observed as spots. Note also that, in Figure~\ref{fig1}d, a few of these spots appear much brighter and correspond to Pb atoms coordinated between two PTO molecules (i.e. Pb$^{2+}$). Since Pb(II) is diamagnetic, they do not show any spin excitations as evidenced in the d$I$/d$V$ spectra of SI Figure S9c. Figures~\ref{fig6}c shows the d$I$/d$V$ cross-section along a Fe-PTO chain (dashed line of Figure~\ref{fig6}a) using a superconducting Pb-tip at 1 K. Figure~\ref{fig6}b is a waterfall plot of the same dataset. $i$, $ii$ and $iii$ refers to the position of Fe atoms in both Figures~\ref{fig6}b and c.  In d$I$/d$V$ spectra on bare Pb(111), the depletion of the density of states around $E_{\textrm{F}}$ due to the superconducting state of the surface ($\Delta_{\textrm{sample}}$) and the tip ($\Delta_{\textrm{tip}}$) is observed as a gap separated by two sharp coherence peaks located at 2$\Delta$ = $\pm$ 2.7 meV.

On PTO molecules, the position and lineshape of the superconducting gap $\Delta$ are identical to the bare Pb(111) (blue curve of Figure~\ref{fig6}d). This indicates that the superconducting state is unaffected by the presence of the PTO precursor implying the absence of any magnetic interaction from the molecule. At Fe sites (red curve in Figure~\ref{fig6}b), d$I$/d$V$ spectra shows an asymmetric superconducting gap framed by two pairs of resonances at $\pm\varepsilon_{1}^*$ = $\pm$3.9 meV and $\pm\varepsilon_{2}^*$ = $\pm$4.7~meV outside the gap. In analogy to the Fe-PTO chain on Ag(111), we assigned these features to the series of low-energy spin excitations of the Fe(II)-PTO complex on Pb(111). Comparing spectra at the $i$, $ii$ and $iii$ locations further shows slight variations in energy of these resonances (SI Figure~S6 and Table~S1), which we attribute to a variation of the coupling of Fe spins with the Pb substrate.\cite{Franke2011,Abdurakhmanova2013} 
\begin{table}[t!]
	\begin{center}
	\caption{Comparison of ($D$,$E$) parameters and spin-excitation energies $\lvert\varepsilon_{n}\rvert$ in meV of the Fe-PTO chain adsorbed on Ag(111) and on Pb(111) extracted from simulations.}\label{excitation energies}
		\begin{tabular}{c c c}
\toprule
\multirow{2}{1em}{} & \multirow{2}{7em}{on Ag(111)} & \multirow{2}{7em}{on Pb(111)}\\[2.5ex]
\midrule
($D$,$E$) & (-0.8, 0.25) & (-0.51, 0.1)  \\
$\lvert\varepsilon_{0}\rvert$ & 0.20 & 0.12  \\
$\lvert\varepsilon_{1}\rvert$ & 1.90 & 1.24 \\
$\lvert\varepsilon_{2}\rvert$ & 3.30 & 2.07\\
 $\lvert\varepsilon_{\textrm{Fe-Fe}}\rvert$& 20 & -\\
\bottomrule
		\end{tabular}
	\end{center}
\end{table}
Coordinated Fe adatoms on Pb(111) are in a $S$ = 2 high spin state. On superconductor, the energies of spin excitations in d$I$/d$V$ spectra are symmetric with respect to $E_{\textrm{F}}$ but shifted by $\pm$2$\Delta$ = $\pm$ 2.7 meV due to the superconducting state of the tip and the substrate.\cite{Berggren2015,Berggren2015a} Experimental resonances $\pm\varepsilon_{1}^*$ and $\pm\varepsilon_{2}^*$ on Pb(111) thus correspond to spin excitation energies $\lvert\varepsilon_{1}\rvert$ = $\lvert\varepsilon_{1}^*\rvert$ - $2\Delta$ =  1.2 meV and $\lvert\varepsilon_{2}\rvert$ = $\lvert\varepsilon_{2}^*\rvert$ - $2\Delta$ = 2.0 meV, respectively. They do not match the ones observed on Ag(111) as shown in the Table~\ref{excitation energies}. 

To explain this discrepancy, we next reproduce d$I$/d$V$ spectra for a single Fe(II) impurity on a metal surface using the phenomenological spin hamiltonian with the parameters $D$ = -0.51 meV and $E$ = 0.1 meV, $U$ = 0.01~meV $J_K\rho_s$= -0.27~meV (SI Figure S7a). This spectrum was then shifted by $\pm$2$\Delta$ at each side of the Fermi level as schematically shown in SI Figure S7b, which now provides a good agreement with the experimental data of Figure~\ref{fig6}d. The excitation energies obtained from simulation are $\varepsilon_0$ = 0.12 meV, $\varepsilon_1$ =  1.24 meV, and  $\varepsilon_2$ =  2.07 meV, respectively. The $\varepsilon_{0}$ energy on Pb(111) thus emerges at $\varepsilon_{0}^*$ $\approx$  $\pm$2.8 meV near the superconducting gap edges. This explains the small dips in d$I$/d$V$ spectra after the gap edges on Fe atoms ($\pm\varepsilon_{0}^*$ in Figure~\ref{fig6}d), which are absent on PTO molecules or Pb(111). 

Figure~\ref{fig6}e shows d\textit{I}/d\textit{V} maps acquired at energies $\pm\varepsilon_0^*$, $\pm\varepsilon_1^*$ and $\pm \varepsilon_2^*$, respectively. For the $\varepsilon_1^*$ and $\varepsilon_2^*$, the DOS is more pronounced at Fe sites while PTO molecules show a homogeneous background, which indicates the localization of the spin-flip process at the Fe atoms. Note also that small variation in intensity at Fe atoms can be observed, which is due to the slight shift in energy of these spin excitations as a function of the position (Figure~\ref{fig6}b). The map at $\pm\varepsilon_0^*$ mainly shows the distribution of DOS at PTO molecules with a darker contrast at Fe sites, thus reflecting the dips of d$I$/d$V$ spectra recorded at Fe atoms (Figure~\ref{fig6}d). 

We last applied a magnetic field of 0.5 T perpendicular to the sample in order to quench the superconducting state of both tip and substrate (SI Figure~S10). By inducing such a metallic state, the spin excitation energies $\varepsilon_n$ of Fe are not shifted anymore by the tip/sample superconducting gaps and now appears near $E_{\textrm{F}}$. Similar to Ag(111), the experimental d$I$/d$V$ spectra thus shows a dip centered at $E_{\textrm{F}}$ (red spectra, SI Figure S10d) which is not observed on  bare Pb(111) (blue spectra, SI Figure S10d).

\subsection*{Discussion}
In contrast to Ag(111), spin-spin interactions between neighboring Fe atoms expected at large energy was never observed on Pb(111) with and without an external magnetic field (SI Figure S11).  As reported in Reference,\cite{Abdurakhmanova2013} this difference between the two substrates is likely governed by their band structures. The Ag(111) surface exhibits a Shockley surface state at about -68~meV with a parabolic dispersion close to the $\Gamma$-point,\cite{Burgi2000} while Pb(111) has a flat dispersion around $M$- and $K$-point at -2.2~eV.\cite{Wurde1994} We think that the superexchange interaction between Fe atoms depends on the amount of charge transfer from the substrate to the Fe-PTO system. The strong hybridization of the molecule with the Ag(111) states might vary the charge distribution of Fe(II)-PTO bonding and the strength of the superexchange mechanism (SI Figures S3 and S4), leading to a strong Fe-Fe ferromagnetic coupling as opposed to the Pb(111) case.
 
In conclusion, we used coordination chemistry to synthesize an iron-based metal-organic ladder using PTO precursors as ligands. We compared the magnetic signature of the high-spin Fe(II) using tunneling spectroscopy at low temperature. The iron coordination with PTO decouples the Fe(II) magnetic moment from both substrate,\cite{Franke2011,Abdurakhmanova2013}  allowing the signature of spin excitations in d$I$/d$V$ spectra. A strong magnetic anisotropy localized at Fe sites is induced by the Ag(111) surface and accompanied with a long-range spin-spin coupling due to superexchange interaction through the PTO ligands. In contrast, the spin-spin interaction is suppressed on Pb(111). The difference of the superexchange interaction between the two substrates is attributed to the depletion of electronic states on Pb(111) as compared to Ag(111), which likely decrease the hybridization of the PTO precursors with the surface and the amount of charge transfer to the Fe-PTO bonding. Overall, our results demonstrate the synthesis of a two-dimensional metal-organic structures on a superconductor that will motivate the design and fabrication of low-dimensional 
magnetic superconductor hybrids with potentially protected spin states.

\section*{Acknowledgements}
Financial support from the Swiss National Science Foundation (SNSF grant 200021\_204053, 200021\_228403 and CRSII5\_213533). The Swiss Nanoscience Institute (SNI) and supports as a part of NCCR SPIN, a National Centre of Competence (or Excellence) in Research, funded by the SNF (grant number 51NF40-180604)  are gratefully acknowledged. We also thank the European Research Council (ERC) under the European Union Horizon 2020 research and innovation programme (ULTRADISS Grant Agreement No. 834402). This work is under the scope of the QUSTEC program, which has received funding from the European Union's Horizon 2020 research and innovation program under the Marie Sk{\l}odowska-Curie grant number 847471. C.L. acknowledges the Georg H. Endress Foundation for financial support. DFT calculations were financially supported by the SNSF Professorship PP00P2\_187185 and performed on UBELIX (http://www.id.unibe.ch/hpc), the HPC cluster at the University of Bern.

\section*{Conflict of Interest}
The authors declare no conflict of interest.

\begin{shaded}
\noindent\textsf{\textbf{Keywords:} \keywords} 
\end{shaded}

\setlength{\bibsep}{0.0cm}
\bibliographystyle{Wiley-chemistry}
\bibliography{Manuscript}

\clearpage
\begin{figure*}[t!]
\begin{center}
\includegraphics[width=5.0cm]{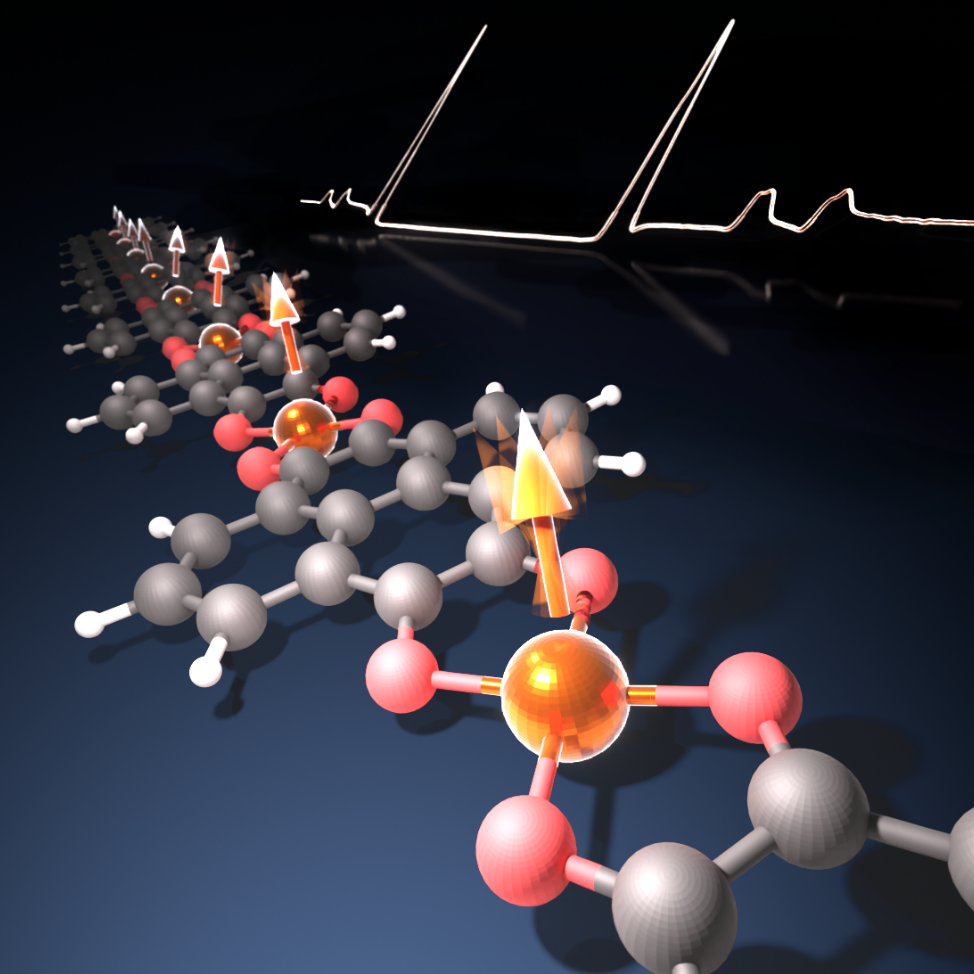}
\caption{Table of content image.\\ 
A spin chain is formed with metal-organic frameworks on surfaces of Ag(111) and Pb(111). Tunneling spectroscopy shows spin-flip excitations from high spin Fe(II) as well as a long-range spin-spin coupling on Ag(111), which is absent on Pb(111) due to the depletion of electronic states near the Fermi level. 
}
\label{TOC}
\end{center}
\end{figure*}
\end{document}